\documentstyle[11pt]{article}
\begin{document}
\parskip=5pt plus 1pt minus 1pt

\vspace{0.5cm}

\begin{center}
{\Large \bf Global monopoles in the Brans-Dicke theory}
\end{center}

\vspace{0.4cm}

\begin{center}
{\bf Xin-zhou Li} \footnote{e-mail address: {\it xzhli@ecust.edu.cn}}
and {\bf Jizong Lu} \footnote{e-mail address: {\it jzlu@shtu.edu.cn}}  \\
{\sl Department of Physics, Shanghai Normal University, Shanghai 200234,
P.R. China} \\
\end{center}

\vspace{3cm}

\begin{abstract}
A gravitating global monopole produces a repulsive gravitational field
outside the core in addition to a solid angular deficit in the Brans-Dicke
theory. As a new feature, the angular deficit is dependent on the values
of $\phi_{\infty}$ and $\omega$, where $\phi_{\infty}$ is asymptotic
value of scalar field in space-like infinity and $\omega$ is
the Brans-Dicke parameter.

\medskip

\begin{center}
(PACS numbers: 04.90.+e, 14.80.Hv, 95.30.Sf,)\\

\medskip

\end{center}

\end{abstract}

\newpage

\vspace{0.4cm}

The phase transitions in the early Universe can give rise to topological
defects of various kinds \cite{1}. The idea that monopole ought to exist
has proved to be remarkably durable. The first ones to study the effects
of gravity on the global monopole were Barriola and Vilenkin \cite{2}.
When gravity is taken account the linearly divergent mass of a global
momopole has an effect analogous to that of a tiny mass at the origin.
Harari and Loust\`{o} \cite{3}, and Shi and Li \cite{4} have shown that
this small gravitational potential is actually repulsive. Recently, Li
{\it et.al.} \cite{5,6,7}have described a new class of cold stars, which
are called D-stars (defect stars). Comparing to Q-stars, one further
requires, as a new feature, that in the absence of matter field the theory
has monopole solutions. This requirement is such that the characteristics
of these objects, for instance deficit angle, differ quite substantially
from those of Q-starts. On the other hand, there has been renewed
interest in the Brans-Dicke (BD) theory \cite{8}, in which the usual metric
gravitational field is augmented by a scalar field $\phi$ which couples
to the curvature via a parameter $\omega$. The modern studies of BD theory
are motivated by the fact that they appear as the low energy limit of string
theory \cite{9,10}. Spherically symmetric charge distributions in BD theory
have also been investigated before \cite{11,12,13,14}. In this paper, we
study global monopoles in the BD theory. We show that the monopole produces
a repulsive granvitational field outside the core in addition to a solid
angular deficit. As a new feature, the angular deficit is dependent on
the values of $\phi_{\infty}$ and $\omega$, where $\phi_{\infty}$ is
asymptotic value of scalar field in space-like infinity.
 
\vspace{0.2cm}

To be specific, we shall work within a particular model in units $G=C=1$,
where a global $O(3)$ symmetry is broken down to $U(1)$ in the framework
of BD theory. Its action is given by
$$
{\cal S} = {1\over{16\pi}}\int d^4 x\sqrt{-g}\,[\phi R - {\omega\over{\phi}}
g^{\mu\nu}\phi,_{\mu}\phi,_{\nu} - 8\pi g^{\mu\nu}\psi^a,_{\mu}\psi^a,_{\nu}
- 4\pi\lambda(\psi^a\psi^a - \eta^2)^2]  \eqno(1)
$$
where $g=\det (g_{\mu\nu})$. $R$ is the scalar curvature, $\omega$ is
a constant, $\phi^a$ is a triplet of the Coldstone field, $a=1,2,3$.
The Goldstone field configuration describing a global monopole is
$${\psi}^a = \eta \sigma(r) {x^a\over r} \;\;\;\;\;\;\; {\rm with}\;
x^ax^a = r^2    \eqno(2)
$$
so that we will actually have a monopole solution if $\sigma\rightarrow 1$
at spatial infinity. The general static metric with spherical symmetry
can be written as
$$
ds^2 = -B(\rho) + A(\rho)dr^2 + \rho^2(d\theta^2 +\sin^2\theta d\phi^2)
\eqno(3)
$$
with the usual relation between the spherical coordinates $\rho,\theta,\phi$
and the ``Cartesian" coordinates $x^a$. Let us now introduce a dimensionless
parameter $r\equiv \eta\rho$. From the action (1) and the definition for
$\sigma$, the equations of motion follows:
$$
{1\over A}\,\sigma'' + [{2\over{Ar}} + {1\over{2B}}({B\over A})']\,\sigma'
- {2\over r^2}\,\sigma - \lambda (\sigma^2 - 1 )\,\sigma = 0,   \eqno(4)
$$
where the prime denoting differentiation with respect to $r$. Varying the
action with respect to $g^{\mu\nu}$ and $\phi$ we obtain the field equations
$$
R_{\mu\nu} - {1\over 2}\,g_{\mu\nu} R = {{8\pi}\over\phi}\,T_{\mu\nu}
+ {\omega\over{\phi^2}} (\phi,_\mu \phi,_\nu - {1\over 2} g_{\mu\nu}
\phi^{,\alpha} \phi,_{\alpha}) + {1\over\phi} (\phi,_{\mu;\nu}
- g_{\mu\nu} \Box\phi)  \eqno(5)
$$
and
$$
\Box\phi = {{8\pi T}\over{2\omega +3}},   \eqno(6)
$$
where the energy-momentum tensor is
$$T_{\mu\nu} = \partial_\nu \psi^a \partial_{\nu}\psi^a
- {1\over 2} g_{\mu\nu}
[g^{\alpha\beta} \partial_\alpha \psi^a \partial_\beta \psi^a
+ {1\over 2} \lambda (\psi^a\psi^a - \eta^2)^2]   \eqno(7)
$$
and $T$ is ${\rm tr}\,T_{\mu\nu}$. Using Eqs. (2) and (3), we have
$$
A' = Ar[{{4\pi\eta^2\sigma'^2(2\omega+1)}\over{\phi(2\omega+3)}}
+ {{8\pi\eta^2A(2\omega-1)}\over{\phi(2\omega+3)}} ({{\sigma^2}\over{r^2}}
+ {\lambda\over 4} (\sigma^2-1)^2) - {{B'\phi'}\over{2B\phi}}
- {{\omega\phi'^2}\over{2\phi}} + {1\over r^2} (1-A) ],   \eqno(8)
$$
$$
B' = {{rB}\over{2\phi+r\phi'}} [ 8\pi\eta^2\sigma'^2
- 8\pi\eta^2A ({\sigma^2\over r^2} + {\lambda\over 4}(\sigma^2-1)^2)
- {4\over r}\phi' + {\omega\over \phi}\phi'^2 + {{2\phi}\over r^2} (A-1)],
\eqno(9)
$$
$$
\phi'' = \phi' ({A'\over{2A}} - {B'\over{2B}} - {2\over r})
- {{2A\eta^2}\over{2\omega + 3}} [{{2\sigma^2}\over r^2}
+ {\sigma'^2\over A} + \lambda(\sigma^2-1)^2].    \eqno(10)
$$
Eqs. (8)-(10) can be reduced to the general relativistic (GR) ones
\cite{2,3,4} when $\omega\rightarrow\infty$ and $\phi\rightarrow 1$.
We introducr the ``scalar charge" of BD theory
$$
S = \lim_{r\rightarrow\infty}\,(r^2{{d\phi}\over{dr}}).   \eqno(11)
$$
Note that this is not in general a conserved quantity.
We define it here since it is included in the expression for the mass of
the monopole which we will give below. The equation for $\phi$ can be
formally intergrated, we have an alternative expression
for the scalar charge
$$
S = \int^{\infty}_0 dr\,r^2\{\phi'({A'\over {2A}} - {B'\over{2B}})
- {{2A\eta^2}\over{2\omega + 3}} [{{2\sigma^2}\over r^2}
+ {\sigma'^2\over A} + \lambda(\sigma^2-1)^2]\}.    \eqno(12)
$$
Expanding the metric and the scalar field equation in powers of $r^{-1}$
about $r = \infty$ one can write the field equations in a linearized form.
Using the boundary condition of the asymptotically flat
$$
A \sim A_\infty + O({1\over r^n}),
$$
$$
B \sim B_\infty + O({1\over r^n}),
$$
$$
\phi \sim \phi_\infty + O({1\over r^n}),    \eqno(13)
$$
where $n\geq 1$ and the subscript $``\infty"$ denotes values at spacelike
$\infty$. We also require $\sigma\rightarrow 1$ at least as fast as $r^{-1}$.
In fact, as we will show below, the conditions above imply that
$\sigma\rightarrow 1$ exponentially as $r\rightarrow\infty$.

The equation for the metric coefficient $A(r)$ reads
$$
({r\over A})' + \Omega(r) ({r\over A}) = 1 - \Delta(r) [\sigma^2
+ {\lambda\over 4} r^2 (\sigma^2 - 1)^2],     \eqno(14)
$$
where
$$\Omega(r) = {{4\pi\sigma'^2\eta^2(2\omega+1)r}\over{(2\omega+3)}}
- {{B'\phi'}\over{2B\phi}} + {{\omega\phi'^2}\over{2\phi}}   \eqno(15)
$$
and
$$
\Delta(r) = {{8\pi\eta^2(2\omega-1)}\over{\phi(2\omega+3)}}.    \eqno(16)
$$
We define
$$\Delta \equiv \Delta(\infty)
= {{8\pi\eta^2(2\omega-1)}\over{\phi_\infty(2\omega+3)}}     \eqno(17)
$$
and we will show below, $\Delta$ describes a solid angular deficit
in the BD theory. Integrating Eq.(14), $A(r)$ can be written as
$$
A^{-1} = 1 - \Delta - {{2M(r)}\over r},      \eqno(18)
$$
where
$$M(r) = {1\over 2}[\Delta(r) -\Delta]r + {1\over 2}\exp[-\int^r_0\Omega(y)dy]
\int^r_0dy\{\Delta(y)[(\sigma^2-1) +{{\lambda y^2}\over 4}(\sigma^2-1)^2
$$
$$
+(1-\Delta(y))y\Omega(y)] +
8\pi{\phi'\over\phi}\Delta(y)\}[\int^y_0\Omega(z)dz].    \eqno(19)
$$
One can show that $\lim_{r\rightarrow\infty}M(r)=M_{ADM}$ which is
Arnowitt-Deser-Misner mass\cite{15} of the monopole. Let us first discuss
the GR case. When $\omega\rightarrow \infty$ and $\phi\rightarrow 1$,
we have $\Delta=8\pi G\eta^2$ and
$$
M(r) = {\Delta\over 2}\exp [-{\Delta\over 2}\int^r_0 dy\sigma'^2(y)y]
\int^r_0 dy[(\sigma^2-1) + {y^4\over 4}(\sigma^2-1)^2
+ (1-\Delta){y^2\over 2}\sigma'^2]\exp[{\Delta\over 2}\int^y_0dz\sigma'^2 z],
\eqno(20)
$$
which is known as a formula in Ref. \cite{3}. Analogously,
$$
B(r) = 1 - \Delta - {{2M_B(r)}\over r}  \eqno(21)
$$
with
$$
M_B(r) = M(r)\exp[\Delta\int^r_{\infty} dy\sigma'^2y]
+ (1-\Delta){r\over 2}[1 - \exp(\Delta\int^r_{\infty} dy\sigma'^2y)].
\eqno(22)
$$
One finds the asympototic expansions
$$
f(r) = 1 - {1\over r^2} - {{{3\over 2}-\Delta}\over r^4} + O(r^{-6}),
$$
$$
M(r) = M_{ADM} + {\Delta\over{2r}} + O(r^{-3}),
$$
$$
M_B(r) = M_{ADM}[1-{\Delta\over r^4}] + {{(1-\Delta)}\over 2}{\Delta\over r^3}
+ O(r^{-7}).   \eqno(23)
$$
Solving Eqs. (8)-(10) in a linearized form, we find scalar field
have following asymptotic form
$$
\phi = \phi_{\infty} - {S\over r} + O({1\over r^2}),   \eqno(24)
$$
while the line element in this limit is
$$
ds^2 = -[1-\Delta-{{2M_K}\over r}+O({1\over r^2})]dt^2
+ [1-\Delta-{{2M_{ADM}}\over r} + O({1\over r^2})]^{-1}dr^2
+ r^2(d\theta^2+\sin^2\theta d\phi),   \eqno(25)
$$
where
$$
M_K = M_{ADM} - {S\over\phi_\infty}   \eqno(26)
$$
is the Keplerian mass of the monopole. The $M_K$ is the active gravitational
mass measureed by a non-self-gravitating test particle in a circular orbit
at space-like infinity about the monopole. Substituting the metric components
appreaing in the asymptotic form of the line element in Eq.(4), we have
$$
(1-\sigma)'' + {2\over r}(1-\sigma)'
- 2\lambda(1+\Delta+{{2M_{ADM}}\over r})(1-\sigma) + O({1\over r^2}) = 0.
\eqno(27)
$$
We obtain the asymptotic solution
$$
\sigma = 1 - r^{-b}e^{-kr}[1+O({1\over r^2})],   \eqno(28)
$$
where
$$
k = [2\lambda(1+\Delta)]^{1\over 2},
$$
$$
b = 1 + ({{2\lambda}\over{(1+\Delta)}})^{1\over 2}M_{ADM}.   \eqno(29)
$$
Therefore, $\sigma$ field tends to $1$ exponentially with $r$
in the asymptotic region.

Neglecting the mass term in Eqs. (18) and (21) and rescaling variables
$r$ and $t$ at large distance off the core, the monopole metric can be
rewritten as
$$
ds^2 = -dt^2 + dr^2 - (1-\Delta)r^2(d\theta^2+\sin^2\theta d\phi^2)
\eqno(30).
$$
This metric describes a space with a deficit solid angla: the area
of a sphere with radius $r$ is not $4\pi r^2$, but a little smaller.
In the BD theory, angular deficit is dependent on the values of
$\phi_\infty$ and $\omega$.

Note that the integrand in Eq. (19) contains a non-positive definite term,
so $M(r)$ may be negative. The strongest constrains on BD theories are
usually assumed to come from the solar system weak field tests \cite{15}.
As is well known, observations constrain the BD parameter to have a value of
$\omega > 500$. Using numerical integration, $M(r)$ is indeed negative all
the way from the origin and quickly approaches an asymptotic value of order
$M\approx -6\pi\eta^2(2\omega-1)/[\phi_{\infty}(2\omega+3)]$. However,
one has taken $\omega = -1$ in the simplest string effective action \cite{10},
which is relavant to the study of global monopole in very early Universe,
although in this case $\phi_\infty =1$ is physically unrealistic:
to approximate a monopole embedded in the string dominated Universe,
we should choose a value $\phi_\infty < 1$. In this case, we have
$\Delta = -24\eta^2/\phi_\infty$. This means that there is not angular deficit,
but angular surplus in the string dominated cosmology.

Finally, the general scalar-tensor gravitational theories arise
from dimension reduction of higher dimensional theories \cite{17}
and string theory \cite{10}. The general theories can satisfy
the rather strict constraints determined at the current epoch,
while still differing considerably from Einstein theory in the past \cite{18}.
Details of the monopole in more general scalar-tensor theories will
be considered elsewhere.

\newpage

\begin{flushleft}
{\large\bf Acknowledgments}
\end{flushleft}

\vspace{0.4cm}

This work was partially supported by National Nature Science Foundation
of China under Grant No. 19875016, and National Doctor Foundation of
China under Grant No. 1999025110.

\end{document}